\documentclass[doublecol]{epl2}
\usepackage{graphicx}
\usepackage[intlimits]{amsmath}

\title{Interacting resonant level coupled to a Luttinger liquid:
Universality of thermodynamic properties}

\author{M. Goldstein \and Y. Weiss \and R. Berkovits}
\institute{The Minerva Center, Department of Physics, Bar-Ilan
University, Ramat-Gan 52900, Israel}

 {Fermions in reduced dimensions
(anyons, composite fermions, Luttinger liquid, etc.)}

\abstract{%
We investigate a model of a single resonant level
coupled to the edge of a quantum wire in the Luttinger liquid phase
or to the middle of a chiral Luttinger liquid
via both tunneling and a contact interaction.
Utilizing the Yuval-Anderson approach, we map this model onto
a classical 1D Coulomb gas
in which all the details of both the interactions in the lead and
the level-lead interaction enter only through the corresponding
Fermi-edge singularity exponent,
which we explicitly evaluate using the Bethe ansatz solution
for a particular model of the lead.
Thus the population, dynamical capacitance and level entropy
are universal in the sense of being
equal for models with interactions differing in magnitude
and even in sign.
We demonstrate this to hold quantitatively
using density matrix renormalization group calculations.
Since the Coulomb gas description is of the single channel Kondo
type, we infer that the universality we found implies that
Luttinger liquid physics has no qualitative effect on these
properties, in contrast with perturbative results.
}

\begin{document}

\maketitle

\section{Introduction}
Understanding the properties of strongly correlated systems
has been one of the main fields of investigation in condensed
matter physics in recent years. 
An important class of such problems is that of quantum impurities,
i.e., systems with a finite number of degrees of freedom coupled
to reservoirs of non-interacting particles, the best known examples
of which are the Kondo and Anderson models \cite{hewson}
and the spin-boson model \cite{spin_boson}. 
Another important type, for which non Fermi liquid physics is well
established, is that of one dimensional electronic systems.
When no symmetry is spontaneously broken, the low energy
physics of those systems is described by the Luttinger liquid (LL)
theory, where the quasiparticles are bosonic modes of density (or,
in a dual description, phase) fluctuations \cite{bosonization}.
It is then natural to try to bring these two themes together, by
studying quantum impurities coupled to LLs: from the quantum impurity
perspective, the reservoir now has a non-trivial physics of its
own; from the LL point of view, this gives a way to probe
the intricate physics of the electrons which are coupled
to the impurity (and not the relatively simple behavior of the
weakly-interacting bosons).
Besides these fundamental motivations, quantum impurities (e.g.,
quantum dots) and LLs (e.g., quantum wires), are the basic
ingredients of nano-scale circuits, so that understanding them
has a profound importance for applications.
Although such models have
been studied theoretically for some time, most effort has been
concentrated at understanding transport properties
\cite{bosonization,kane92};
other phenomena have usually received only scant attention
\cite{furusaki02,lehur05,sade05,wachter07,weiss08,lerner08,bishara08}. 

\section{Model}
The simplest possible system to study these
effects is that of a single level coupled to the edge of a LL
(which can be realized by, e.g., lithographically defining a
small quantum dot at the end of a quantum wire, or by
coupling a metallic grain to the edge of a metallic nanowire,
or an impurity atom to the end of a carbon nanotube),
or, equivalently, to the middle of a chiral LL (e.g., a dot
near the edge of a fractional quantum hall bar) \cite{chang03}.
We include contact interaction between the
level and the lead. Here transport properties are not relevant;
however, many other interesting questions can be investigated.
In this letter we concentrate on thermodynamic
properties: the level population, its dynamical
capacitance  (which can be probed experimentally by capacitively
coupling the system to a quantum point contact)
and the level contribution to the entropy and specific
heat. We demonstrate, both analytically and numerically, that
these properties show universality, and depend on the different
interactions only through a single parameter, the Fermi
edge singularity exponent of the system. Using an analogy to the
single channel Kondo problem, we show this universality to imply
that these physical quantities have an essentially Fermi liquid like
behavior, which is affected only quantitatively, but not qualitatively,
by LL physics.

The system is described by the following Hamiltonian:
\begin{multline} \label{eqn:hll_cont}
H = H_{\text{lead}} \left\{ \psi^{\dagger}(x), \psi(x) \right\}
+ \varepsilon_0 d^{\dagger}d
- \left[ \gamma_{\text{ll}} d^{\dagger} \psi(0) + \text{H.c.} \right] \\
+ \frac{\lambda_{\text{ll}}}{2}
\left( d^{\dagger}d - {\textstyle\frac{1}{2}} \right)
\left [\psi^{\dagger}(0) \psi(0) - \psi(0) \psi^{\dagger}(0) \right],
\end{multline}
where $d$, $\psi(x)$ are Fermi operators of the level and the lead,
respectively, $\varepsilon_0$ is the bare level energy,
$\gamma_{\text{ll}}$ is the level-lead tunneling matrix element, and
$\lambda_{\text{ll}}$ is the strength of the level-lead interaction.
At low energies the lead Hamiltonian assumes the Tomonaga-Luttinger form.
It can then be written
in terms of two
Bose fields $\Theta(x)$ and $\Phi(x)$ obeying the
commutation relation $[\Theta(x),\Phi(x')]=\text{i}\pi\theta(x-x')$
[$\theta(x)$ is the step function],
and the boundary condition $\Theta(0)=0$
\cite{bosonization}:
\begin{equation} \label{eqn:hlead_cont}
 H_{\text{lead}}= \frac{v}{2\pi}
 \int_{0}^{\infty}
 \left\{ \frac{1}{g} [\partial_x \Theta(x)]^2 + g [\partial_x \Phi(x)]^2  \right\}
 \text{d}x,
\end{equation}
where $g$ and $v$ are the usual LL interaction parameter and
excitation velocity, respectively.
The electron density equals $\partial_x\Phi/\pi$,
and the electron annihilation operator at the the edge of the lead can be
expressed in the bosonic language as
$\psi(0)=\chi \text{e}^{\text{i}\Phi(0)}/ \sqrt{2\pi a}$,
using a Majorana Fermi operators $\chi$ and a short distance cutoff (e.g.,
a lattice spacing) $a$.

\section{Yuval-Anderson approach and universality}
Using the Yuval-Anderson approach \cite{yuval_anderson}, in either the
canonical \cite{si93} or the path-integral \cite{kamenev_gefen}
formulations, we expand the partition function to all orders
in $\gamma_\text{ll}$ and evaluate the resulting terms.
The expression thus obtained is a sum over all possible imaginary time
histories of the level, which fluctuates between the empty and filled states.
We then obtain the expression:
\begin{multline} \label{eqn:cg1}
Z = \sum_{ \substack{N=0 \\ \sigma=\pm1}}^{\infty}
\left( \frac{\Gamma_0 \xi_0}{\pi} \right)^N
\int_0^{\beta} \frac{d\tau_{2N}}{\xi_0}
\int_0^{\tau_{2N}-\xi_0} \frac{d\tau_{2N-1}}{\xi_0}
\dots \\
\int_0^{\tau_3-\xi_0} \frac{d\tau_2}{\xi_0}
\int_0^{\tau_2-\xi_0} \frac{d\tau_1}{\xi_0}
\exp \left[ - \mathcal{S} (\{ \tau_i \}, \sigma) \right] ,
\end{multline}
where $\xi_0$ is a short time (ultraviolet) cutoff, $\Gamma_0$ is the
(renormalized) level width (an expression for which is given below),
and $\beta$ is the inverse temperature of the original problem.
This expression thus has the form of a classical grand canonical
partition function of a one dimensional gas of particles
(``Coulomb gas'') residing
on a circle of circumference $\beta$, with fugacity
$\sqrt{\Gamma_0\xi_0/\pi}$. Each particle is assigned a positive
(negative) charge if it corresponds to hopping of an electron from
the lead to the level (vice-versa).
The charges must thus be alternating,
with an overall charge neutrality. Hence, a configuration is
completely specified by the sign of the first charge (denoted by
$\sigma$ in the above expression) and by the positions of the particles.
The action of this classical system consists of two terms:
\begin{multline} \label{eqn:cg2}
\mathcal{S}( \{ \tau_i \}, \sigma ) =
\alpha_{\text{FES}}
\negthickspace \negthickspace \negthickspace 
\sum_{1 \le i<j \le 2N} \negthickspace \negthickspace \negthickspace
(-1)^{i+j}
\ln \left\{ \frac{ \pi \xi_0 / \beta }
{\sin [ \pi(\tau_j-\tau_i)/\beta ] } \right\} \\
+ \varepsilon_0 \left[ \beta\frac{1-\sigma}{2} +
\sigma \sum_{1 \le i \le 2N} (-1)^i \tau_i \right].
\end{multline}
The first term is an interaction between the particles, with the
form of a Coulomb interaction between charged rods, and a coefficient
(charge squared) $\alpha_{\text{FES}}$, the Fermi edge singularity
exponent of our problem (by which we refer to twice the scaling dimension
of $d^{\dagger}\psi(0)$ for $\lambda_\text{ll}=0$).
We discuss its value below.
If the lead has a finite length $L$ but the temperature is zero,
one should substitute $L/(\text{i} v)$ for $\beta$ inside the logarithm,
whereas at finite temperature the sine is replaced by
an elliptic function \cite{cft}.
The second term in the action of the classical system
corresponds to the energetic cost of filling the level,
and resembles the effect of an electric field applied on the charges.

We thus see that the partition function depends on the original model
only through three parameters: $\Gamma_0$, $\varepsilon_0$ and
$\alpha_{\text{FES}}$.
As we show below, the latter, in particular, contains the main
effects of the interactions, both in the lead and between the level and
the lead.
This implies a \emph{universality} in this system. We use
this term here to refer to the fact that many of the properties of the system
depend only on these three parameters, so that they will be
the same for very different systems, with different strengths and signs
of interactions, provided these three parameters are indeed the same.
The properties which exhibit
universality are the thermodynamic ones, e.g.: the
level population and its correlation functions (or, equivalently,
the static and dynamic level capacitance),
and the level contributions to the entropy and the specific heat.

Since the universality is based on the Yuval-Anderson
description, it is important to understand the limitations of the latter.
The derivation of the Coulomb gas represtation assumes that
the correlation functions of the tunneling term $d^{\dagger}\psi(0)$
behave as power-laws in time.
While this is correct for the Tomonaga-Luttinger
Hamiltonian~(\ref{eqn:hlead_cont}), any particular model of
a one-dimensional lead will differ from it by terms which are
irrelevant at low-energies (or, equivalently, long times)
in the renormalization group sense.
This will affect the correlation functions in two ways:
(i) At long times they will retain the power-law form, but with
renormalized power and prefactor.
(ii) At short times the the power-law form itself could be modified.
The first effect does not change the form of the Coulomb gas expansion
(or the resulting universality), and can be accounted for by using
the appropriate renormalized values of the Coulomb gas parameters
$\alpha_{\text{FES}}$ and $\Gamma_0$. These values are discussed
in the following section.
The second effect, on the other hand, could have resulted in a
real limitation of the Anderson-Yuval description.
However, numerical data presented below shows that, to a very high
degree of accuracy, this has no quantitative effect,
except in the vicinity of the transition points from the LL phase
to non-metallic phases (where perturbations to the Tomonaga-Luttinger
Hamiltonian become relevant).

\begin{largetable}
\caption{\label{tbl:cg_params}
Parameters appearing in the Coulomb gas model,
eqs.~(\ref{eqn:cg1}) and (\ref{eqn:cg2}).
See the text for further details.
}
\begin{tabular}{|c|c|c|c|}
\hline
& Non-interacting lead & Bosonization & General model \\ \hline \hline
$\alpha_{\text{FES}}$ & $\left( 1 - \frac{2}{\pi} \delta \right)^2$ &
$\frac{1}{g}(1-\frac{g\lambda_{\text{ll}}}{\pi v})^2$ &
$\frac{1}{g} \left( 1 - g\frac{2}{\pi}\delta_{\text{eff}} \right)^2$ \\ \hline
$\Gamma_0$ & $\pi \left| \gamma_{\text{ll}} \right|^2 \nu_0 \cos(\delta)$ &
$\pi \left| \gamma_{\text{ll}} \right|^2 \nu_0$ &
$\pi \left| \gamma_{\text{ll}} \right|^2 \nu_0 \cos(\delta_{\text{eff}})$ \\ \hline
\end{tabular}
\end{largetable}

\section{Coulomb gas parameters}
Following these comments, we now discuss in more details the
parameters appearing in the Coulomb gas form of the partition function.
In the case where the lead is noninteracting, we have the
usual resonant level model,
for which it is known that
$\alpha_{\text{FES}} = \left( 1 - \frac{2}{\pi} \delta \right)^2$ and
$\Gamma_0 = \pi \left| \gamma_{\text{ll}} \right|^2 \nu_0 \cos(\delta)$,
where $\delta=\tan^{-1} (\pi \nu_0 \lambda_{\text{ll}}/2)$ is the phase
shift of the electrons in the lead caused by their interaction with the
level, and $\nu_0$ is the local density of states at the end of the lead
\cite{noziers69,fabrizio95}.
When there are nonvanishing interactions both in the lead and between the
level and the lead, the situation is more complicated. From bosonization
\cite{bosonization} we obtain
(since there is no backscattering in this problem)
$\alpha_{\text{FES}}=(1-g\lambda_{\text{ll}}/\pi v)^2/g$,
and $\Gamma_0=\pi \left| \gamma_{\text{ll}} \right|^2 \nu_0$. If we go to 
the limit of a noninteracting lead
(with $\nu_0=1/(\pi v)$ taking into account both the left-
and right-going branches)%
, we see that in the bosonization treatment
expressions that should contain the phase shift $\delta$ are replaced
by their leading order dependence on $\lambda_{\text{ll}}$.
This is the result of irrelevant corrections to the LL
Hamiltonian~(\ref{eqn:hlead_cont}), in this case --- the band curvature.

On the other hand, from boundary conformal field theory arguments
\cite{affleck94} it follows that $\pi v \alpha_{\text{FES}}/L$ is equal
to the $1/L$ correction to the difference between the two energies:
the energy of a lead with no attached level but with potentials
of strengths $\pm \lambda_\text{ll}/2$ applied on
its two edges, and
the energy of that lead with one electron extracted and a potential of strength
$\lambda_\text{ll}/2$ applied on both ends. It may thus be calculated
numerically, or even analytically when an exact solution is available.
Let us consider, for example, a discrete realization of the lead as a
half-filled tight-binding chain with nearest-neighbor interactions:
\begin{equation} \label{eqn:hlead_disc}
 H_{\text{TBl}} =
   \sum_{i=1}^\infty 
   \left[ -t c_{i}^{\dagger} c_{i+1} +  \text{H.c.}
   + U \left( n_{i} - {\textstyle\frac{1}{2}} \right)
     \left( n_{i+1} - {\textstyle\frac{1}{2}} \right) \right],
\end{equation}
where $c_i$ is the Fermi operator at the lead's $i$th site,
$n_i = c^{\dagger}_{i} c_i$ is the corresponding number operator,
while $t$ and $U$ are the nearest-neighbor
hopping and interaction strengths along the chain.
The LL parameters of this model are $g=\pi/[2\cos^{-1}(-\Delta)]$,
and $v/(t a)=\pi \sqrt{1-\Delta^2} / \cos^{-1}\Delta$, with
$\Delta \equiv U/2t$ and $a$ denoting the lattice spacing \cite{bosonization}.
The full Hamiltonian (including the level) is now:
\begin{eqnarray} \label{eqn:hll_disc}
 H_\text{TB} & = & H_\text{TBl} + \varepsilon_0 d^\dagger d
   - \left( t_{\text{ll}} c_1^{\dagger} d + \text{H.c.} \right) 
   + \\ \nonumber & & U_{\text{ll}}
     \left( d^{\dagger} d - {\textstyle\frac{1}{2}} \right)
     \left( c_{1}^{\dagger} c_{1} -{\textstyle\frac{1}{2}} \right),
\end{eqnarray}
where the level-lead couplings are related to their continuum
counterparts by
${\gamma}_{\text{ll}} = t_{\text{ll}} \sqrt{a}$, and
${\lambda}_{\text{ll}} = U_{\text{ll}} a$. 
This model of the lead (or its equivalent, the XXZ spin chain
\cite{bosonization}) is exactly solvable by the Bethe ansatz
even for a finite size system and in the presence of
potentials at the boundary \cite{woynarovich87,hamer87}. 
Hence, an analytic expression for $\alpha_{\text{FES}}$ can
be found in this case: 
\begin{equation} \label{eqn:alpha_bethe}
\alpha_{\text{FES}} =
\frac{1}{g} \left[ 1 - \frac{2g}{\pi}
\tan^{-1} \left( \frac{ U_{\text{ll}} }
{ \sqrt{(2t)^2-U^2} } \right) \right]^2.
\end{equation}
It then seems natural to identify
$\delta_{\text{eff}}=\tan^{-1} \left( \frac{ U_{\text{ll}} }
{ \sqrt{(2t)^2-U^2} } \right)$
as an effective phase shift, which reduces to the usual phase shift
when the lead is noninteracting.
We may thus expect that for a general model we can write
$\alpha_{\text{FES}} =
\frac{1}{g} \left( 1 - \frac{2g}{\pi}\delta_{\text{eff}} \right)^2$,
for some effective phase shift $\delta_{\text{eff}} \in [-\pi/2,\pi/2]$,
so that $\Gamma_0$ will be given by
$\pi \left| \gamma_{\text{ll}} \right|^2 \nu_0 \cos(\delta_{\text{eff}})$.
In the following we will confirm these results
quantitatively by our numerical data. 
The discussion in the last two paragraphs is summarized
in table~\ref{tbl:cg_params}. 


We note in passing that the mapping into the Coulomb gas can be easily
extended to include the case of an Ohmic environment coupled to the level.
The only effect of this on the analysis is modifying the parameter
$\alpha_{\text{FES}}$ by adding to it the impedance of the environment
divided by the quantum resistance $h/e^2$ \cite{lehur05}.
Hence, all our results apply to this case too.
The universality is thus seen to have an even broader scope of
applicability.

\begin{figure}
\includegraphics[width=8cm,height=!]{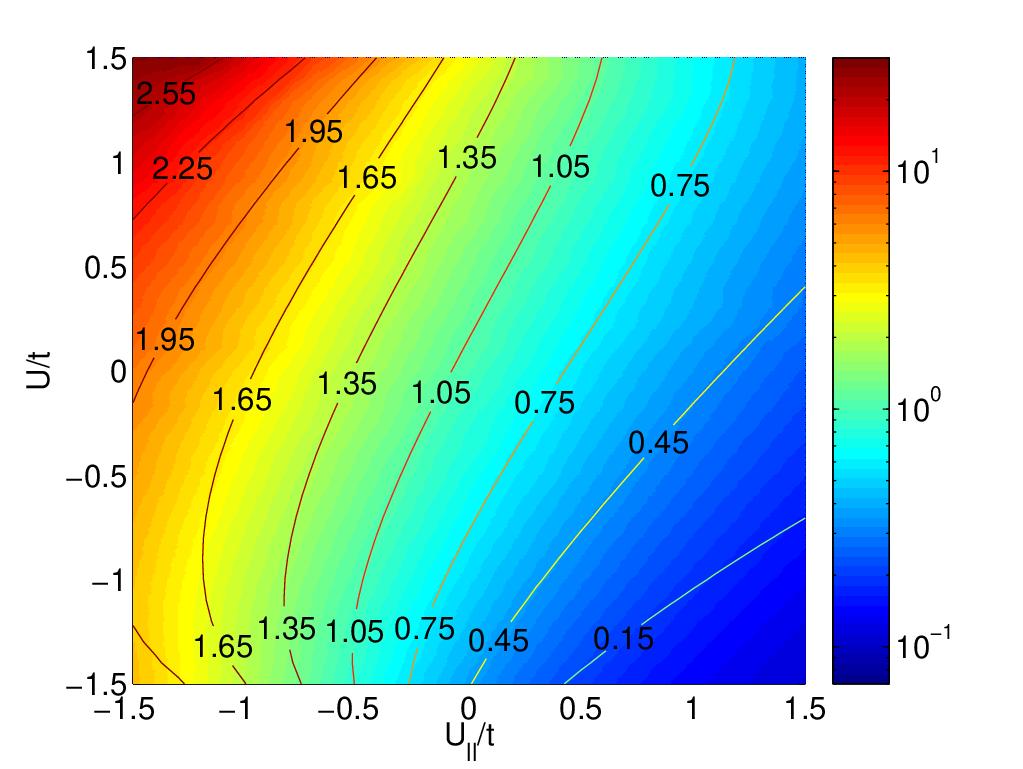} 
\caption{\label{fig:clrmap}
(Color online)
A color map of the differential capacitance
(divided by the noninteracting value)
obtained by DMRG at half filling,
as a function of the interactions in the lead and
between the level and the lead, with
contours of constant $\alpha_{\text{FES}}$ superimposed.
It can be seen that the effect of all the interactions comes
only through this parameter, confirming universality.
See the text for further details.
}
\end{figure}

\section{Numerical results}
Let us now turn to a numerical test of the universality.
As we explained above, this enables us to show that although
the mapping to a Coulomb-gas applies rigorously only to the low
frequency (long time) behavior,
we have found that universality holds \emph{quantitatively},
at least when irrelevant perturbations of the LL are not
too strong. To this end we have performed density matrix renormalization
group (DMRG) \cite{white93}
calculations on the half-filled tight binding realizations of the system,
eqs.~(\ref{eqn:hlead_disc}) and~(\ref{eqn:hll_disc}).
Up to 256 block states were kept in each iteration.
In fig.~\ref{fig:clrmap} we show the differential capacitance
$\partial n / \partial \varepsilon_0$ at $\varepsilon_0=0$ in a color map
as a function of both the level-lead interaction and the interaction in the lead.
In all cases we have kept $\Gamma_0=10^{-4}t$ and $L=50v/t$,
modifying $t_{\text{ll}}$ and $L$ accordingly, so as to keep all
the parameters of the Coulomb gas constant except $\alpha_{\text{FES}}$.
On the color map we superimposed a contour plot of $\alpha_{\text{FES}}$,
taken from eq.~(\ref{eqn:alpha_bethe}). It is indeed seen that
the contours of constant $\alpha_{\text{FES}}$ are also contours of
constant differential capacitance, confirming the important role of the
former in determining the behavior of the system. Deviations are seen only
for quite strong interactions, where irrelevant terms in the Hamiltonian
are initially quite strong (and are not renormalized to zero because
of the finite system size), and thus modify the results quantitatively.
To appreciate this one should remember that for $|U|>2t$
the system is no longer a LL [but becomes charge density wave (phase
separated) for positive (negative) $U$]; whereas for
$|U_{\text{ll}}|>2t$ the potential of
$\pm U_{\text{ll}}/2$ felt at the last site of the
lead when the level is full (empty) is strong enough to form a bound
state. Both of these effects are not included in our treatment.

\begin{figure}
\includegraphics[width=8cm,height=!]{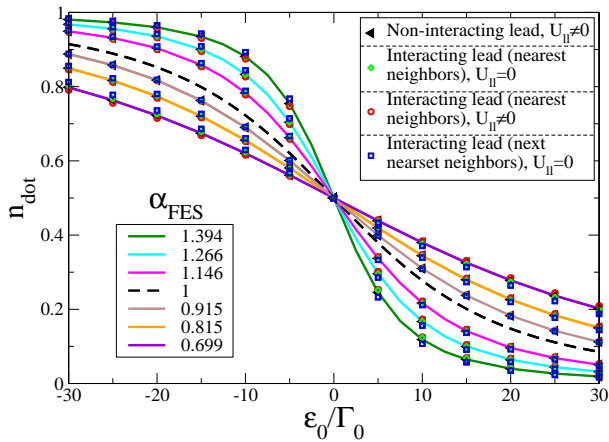}
\caption{\label{fig:dl_nn_nnn}
(Color online)
Level population as a function of its energy: different symbols denote
four models used in the DMRG calculations, while the different
curves (which are a guide to the eye) correspond to different
$\alpha_{\text{FES}}$ values (the smaller $\alpha_{\text{FES}}$
the wider the curve and vice versa).
In the last model the strengths of the nearest neighbor and next nearest
neighbor interactions are:
$\{U/t, V/t\} =
\{1.5, 0.5\}$, $\{1.0, 0.5\}$, $\{0.5, 0.5\}$, $\{-0.25, -0.25\}$,
$\{-0.5, -0.5\}$, $\{-0.75, -0.5\}$,
in order of decreasing $\alpha_{\text{FES}}$.
In the third model $U$ was taken as $\pm 0.5t$,
with \emph{opposite} sign to the corresponding fourth model case.
See the text for further details.
}
\end{figure}

A more detailed comparison is made in fig.~\ref{fig:dl_nn_nnn}.
Here we show the full dependence of the level population on its energy.
The population curves corresponding to different
$\alpha_{\text{FES}}$ values are presented,  and on each
such curve there are symbols of four types, denoting the numerical
results on four different models:
(a) A non-interacting lead with nonzero level-lead interaction;
(b) A lead with nearest-neighbor interactions but zero level-lead term;
(c) A system with both nonzero $U_{\text{ll}}$ and nearest-neighbor
interaction in the lead [which serves as a test to
eq.~(\ref{eqn:alpha_bethe}) and the subsequent discussion];
(d) A lead with next-nearest-neighbor
interactions of strength $V$ in addition to the nearset-neighbor interactions
(but vanishing $U_{\text{ll}}$), i.e., with the term
$V \sum_i \left( c_{i}^{\dagger} c_{i} -\frac{1}{2} \right) 
\left( c_{i+2}^{\dagger} c_{i+2} -\frac{1}{2} \right)$
added to eq.~(\ref{eqn:hlead_disc}).
This is used to show that our results
apply even to non-integrable models
(in this system $g$ was determined numerically).
The parameters of the four models were chosen to give the same
$\alpha_{\text{FES}}$ value [i.e., in each case we have chosen
arbitrarily the interactions in the lead in models (c) and (d),
and determined by the above condition all the other interactions.
The other parameters are the same as in fig.~\ref{fig:clrmap},
except for the lead length, which is twice as large here].
Again we can see that the population is universal,
determined by $\alpha_{\text{FES}}$ alone,
and not by the parameters of a specific model.

\section{Lessons from the Kondo effect}
We now discuss another implication of the Coulomb gas mapping.
The Coulomb gas we
have obtained is similar to the original one, derived by Yuval and Anderson
in their treatment of the anisotropic single-channel Kondo model
\cite{yuval_anderson}.
In particular, the level population (minus one half) in our system
is equivalent to the magnetization of the Kondo impurity, the level energy
$\varepsilon_0$ is analogous to a local magnetic field,
$\Gamma_0$ plays the role of $J_{\perp}$, and $\alpha_{\text{FES}}$
is determined by $J_z$.
We can thus immediately import all the known results from the
Kondo problem \cite{hewson}
to the case of a LL lead coupled to a level. 
The system considered can be in one of two phases: a strong coupling
(antiferromagnetic Kondo like) delocalized phase, and a weak coupling
(ferromagnetic Kondo like) localized phase.
At very small values of $\Gamma_0$ the transition
is at $\alpha_{\text{FES}}=2$, whereas for larger $\Gamma_0$ it occurs
for larger values of $\alpha_{\text{FES}}$.
In the localized phase, the low energy physics is that of an effectively
disconnected level, so that its population is discontinuous as a function
of $\varepsilon_0$, and there is a nonvanishing residual entropy at zero
temperature. Similar results regarding this phase, as
well as the phase transition line, were already discussed in
Ref.~\cite{furusaki02}, albeit using different techniques,
and tested numerically by us\cite{weiss08}.
On the other hand, in the delocalized phase, the impurity is well hybridized
with the conduction band, so the level population is analytic in
$\varepsilon_0$. One can write an explicit expression for this dependence using
the Bethe ansatz solution of the Kondo problem \cite{hewson}.
In particular, for small values of $\varepsilon_0$, one has:
\begin{equation} \label{eqn:ne}
n(\varepsilon_0) \sim {\textstyle\frac{1}{2}} - \frac{\varepsilon_0}{\pi T_K}
\end{equation}
with $T_K$ (the effective level width) corresponding to
the ``Kondo temperature''
of the problem, which, for small $\Gamma_0$, is given by:
\begin{equation} \label{eqn:tk}
T_K =
\left( \Gamma_0 \xi_0 \right)^{1/(2-\alpha_{\text{FES}})} / \xi_0,
\end{equation}
and thus reduces to $\Gamma_0$ for vanishing interactions
($\alpha_{\text{FES}}$=1).
Hence, in this phase the population does not show any power law dependence
on $\varepsilon_0$. The only power law appearing is in the formula for $T_K$.
However, the power depends on $\alpha_{\text{FES}}$, and is nontrivial
(i.e., different from unity) even for a Fermi liquid lead if level-lead
interactions are not negligible, or in the presence of dissipation.
The same conclusion applies to other quantities in this phase:
at long time (denoted by $\tau$) the correlation function of the level
population will decay as $(\tau T_K)^{-2}$,
and the entropy and specific heat will go as $1/(\beta T_K)$ for low
enough temperatures.
These results are in fact another
manifestation of the universality property of this system: it implies
that LL physics (with its ubiquitous power law dependences)
\emph{cannot} be manifested through the behavior of any of the
thermodynamic properties, contrary to what one might expect based
on perturbative calculations%
, like those performed (albeit for a different,
two-lead configuration) in Ref.~\cite{wachter07} for the case
$\lambda_{\text{ll}}=0$.
Such calculations, while reproducing eq.~(\ref{eqn:tk}), deviate from
eq.~(\ref{eqn:ne}) if $g$ is sufficiently small.

From eqs.~(\ref{eqn:ne}) and (\ref{eqn:tk}) we see that the population
curve
becomes wider as $\alpha_{\text{FES}}$ becomes smaller and vice-versa,
in agreement with the numerical results shown in
figs.~\ref{fig:clrmap} and \ref{fig:dl_nn_nnn}.
This has a simple interpretation: smaller $\alpha_{\text{FES}}$
corresponds, according to the previous results,
to large $g$ (i.e., attraction in the lead) or positive
$\lambda_{\text{ll}}$. Indeed, when $g$ is larger than 1, the local
density of states at the edge of a LL (or at the middle
of a chiral LL) diverges at the Fermi energy \cite{bosonization}, so
tunneling is enhanced; similarly, for $\lambda_{\text{ll}}>0$ tunneling
is also enhanced by the Mahan exciton effect \cite{noziers69}: 
When the level is empty (full) the adjacent site of the lead tends
to be full (empty) because of the charging interaction,
so transition between these states becomes easier.
In both cases, the population curve should indeed become broader.

\section{Conclusions}
To conclude, we have shown that the thermodynamic properties of
a level coupled to the edge of a LL
are universal for a wide range of models, and are determined
by only few parameters.
These properties follow a single-channel Kondo physics,
and thus are not qualitatively affected by the LL phase of the lead.
This implies that interesting phenomena occurring in quantum impurities
coupled to LLs can be studied on equivalent models with non-interacting
leads, which are much easier to study, both analytically and numerically
(using, e.g., Wilson's numerical renormalization group \cite{nrg}).
A clear signature of the LL phase can be seen when examining
transport-like properties (e.g., the level local density of states).
Alternatively, one could extend the model to include more than one
lead. Both topics will be discussed elsewhere \cite{goldstein08}.


\acknowledgments
We would like to thank Y.\ Gefen and A.\ Schiller for many useful discussions.
M.~G.\ is supported by the Adams Foundation Program of the Israel Academy
of Sciences and Humanities.
Financial support from the Israel Science Foundation (Grant 569/07) is
gratefully acknowledged.

\end{document}